\documentstyle[prb,aps,multicol,epsf]{revtex}

\newcommand{\startlongequation}{
\end{multicols}\vspace*{-3.5ex}{\tiny\noindent
\begin{tabular}[t]{c|} \parbox{0.493\hsize}{~} \\ \hline \end{tabular}} }
 
\newcommand{\stoplongequation}{
{\tiny\hspace*{\fill}
\begin{tabular}[t]{|c}\hline\parbox{0.49\hsize}{~} \\ \end{tabular}}
\vspace*{-2.5ex}\begin{multicols}{2} }

\begin{document}
\widetext
\draft

\title{Faceting 
and stress transfer in the   missing-row reconstruction
of Ir (110)}
\author{Alessio Filippetti$^{1,2}$ and Vincenzo Fiorentini$^{1,3}$}
\address{{\it (1)}\ Istituto Nazionale per la Fisica della Materia 
and Dipartimento di 
 Fisica, Universit\`a di Cagliari, I-09124 Cagliari, Italy\\
{\it (2)}\ Department of Physics, University of California,
Davis, CA 95616, U.S.A.\\
{\it (3)}\ Walter Schottky Institut, Technische Universit\"a{}t M\"u{}nchen,
Am Coulombwall, D-85748 Garching bei M\"u{}nchen, 
Germany}
\date{\today}
\maketitle
\begin{abstract}
 We present ab initio total energy and stress calculations 
for the unreconstructed and (2$\times$1)--missing-row
reconstructed Ir (110) and Rh (110) surfaces. 
We use a model based on ab initio results to show
 that the ($n$$\times$1) reconstruction  is a faceting transition to
 a long-wavelength--corrugated (111)-like surface, and find the 
3$\times$1 structure to be the most stable for Ir in
accordance with experiment. We then use the stress density to
analyze  the stress increase  upon reconstruction, and 
find it to be due to a changed balance of tensile and compressive
contributions in the near-surface region.  
\end{abstract}

 
\begin{multicols}{2}

\noindent
Experimental findings\cite{exp} clearly support 
the  missing-row reconstruction  model 
for the (110) surfaces of the $5d$ metals Ir, Pt
and Au. \cite{due}  The reconstructed
surfaces have been observed  in a variety of
 $n\times 1$ structures (presumably not all   at equilibrium)
 whereby  $n$--$m$ out of $n$
atomic rows of  first-neighbor atoms, oriented in the [1$\overline{1}$0]
direction,  are removed from the $m^{\rm th}$ plane, so that all
planes through the  ($n$+1)$^{\rm th}$ expose rows to the vacuum. 
Calculations support the existence of this
reconstruction pattern for Au \cite{garofalo,hb} and Ir \cite{mrs} 
and its absence for  Ag \cite{hb} and Rh \cite{mrs}.
For Ir, mixed reconstructions  patterns with predominantly 3$\times$1
character have been observed,\cite{exp_ir} while first-principle
calculations\cite{mrs} confirmed that at least the computationally
affordable 2$\times$1 structure is indeed favored over the 
unreconstructed one. On a different line,  surface stress has been
investigated both experimentally\cite{bertel,lehw}  and
theoretically\cite{mrs,feib-ani,fms}  (mostly for Pt, Pd, Rh, and Ir).  
The behavior and role of stress in the reconstruction 
remains to be investigated. Also, the stability of
 $n$$\times$1 structures 
has not yet been supported theoretically from first-principle
(accurate semiempirical studies exist on Au (110) \cite{garofalo}).

Here we address the existence of the 2$\times$1 reconstruction
 by means of ab initio calculations.
We then set up a model to describe $n$$\times$1 
reconstructions, based on ab initio quantities for the
computationally tractable cases.
Finally, we analyze the behavior of surface
stress using a recently-developed tool, the stress 
density.\cite{long}
In the first-principles local-density-functional
calculations 
we use either fully separable norm-conserving (Ir) or ultrasoft (Rh)
pseudopotentials, a plane-wave basis cut off at 40 Ry (Ir) and
30 Ry (Rh), and			
 surface slabs of up to 11  atomic planes, fully relaxing all
 structures. For further details, see Ref. \onlinecite{mrs}.

\paragraph*{Ab initio results -- } 
In Table \ref{t5} we report  surface energy \cite{fm}  surface
stress\cite{nm}, and workfunction of the  unreconstructed Ir and Rh 
(110) surfaces, and
their changes upon  reconstruction. The
 reconstruction  energy
$H^{\rm  rec}$ is the difference between reconstructed and
unreconstructed  surface energy, i.e.  $H^{\rm  rec} < 0$
means favored reconstruction.  This is
found to be the case for   Ir, not for Rh, in agreement with
experiment.\cite{exp,exp_ir}  For Ir we find an energy gain of 0.03
eV/atom,  close to the  $\sim$ 0.05 eV/atom obtained previously 
\cite{garofalo,hb} for Au.

The unreconstructed surfaces are
under a large tensile stress (negative in our
convention), as already customary from previous work on 
transition metals.\cite{mrs,feib-ani,fms}
The stress is appreciably larger for Ir than for Rh 
due to the  influence of  relativistic effects\cite{fms}),
and  highly anisotropic:  A=$\tau_y^{\rm unrec}/\tau_x^{\rm unrec}$ is
1.89 for Ir and 1.61 for Rh  (here $\hat{x}$ = [001] and $\hat{y}$ =
[1$\overline{1}$0]).
The anisotropy is almost entirely due to relaxations (in 
the ideal structure we find A=1.17 and 1.11 for Ir and Rh), in
agreement with  Feibelman's argument \cite{feib-ani}  that the 
first-neighbor bonds having a component along the $[001]$ direction can
shorten upon relaxation, and relieve a large  part of their tensile
stress, whereas first-neighbor bonds forming the
$[1\overline{1}0]$-oriented atomic rows  cannot shrink because of
symmetry constraints, and thus conserve the stored tensile stress.  
Our A-values are similar to 
Feibelman's\cite{feib-ani} for Pt and Pd.  Both A and its change upon
relaxation are greater for the $5d$
elements -- that is, the larger the surface stress, the larger the
portion of stress relieved by the relaxation.

Reconstruction  causes a puzzling behavior of the surface stress
and workfunction.
As a tensile stress is negative by convention,  negative $\Delta\tau$'s
such as those in Table \ref{t5} indicates  an increase in tensile stress.
The puzzling result is thus that the  surface stress is more tensile on
the reconstructed surface than on the unreconstructed one. This
feature is surprising, as the surface stress is mostly stored into the
bonds of the [1$\overline{1}$0]-oriented  rows, and one would expect
it to decrease upon removal of one row out of two. 
 In addition, the workfunction
increases with reconstruction, suggesting a decrease in surface
roughness. This is surprising because reconstruction decreases  the
surface atomic  density and apparently increases the roughness (the
corrugation amplitude is doubled).   (Although this is not our concern
here, we note that stress does not decrease upon   the 2$\times$1
transition, which seems to disagree with arguments \cite{lehw}  to the
effect that stress is the driving force of this transformation. Stress
has been invoked to explain e.g. the mesoscopic patterning of period
$\sim$ 1500 \AA\, recently  observed   \cite{bertel}  on  2$\times$1
Pt  (110): that, however, is a further transformation of the
2$\times$1-reconstructed surface.)

\begin{figure}
\epsfclipon
\epsfxsize=9cm
\narrowtext
\centerline{\epsffile{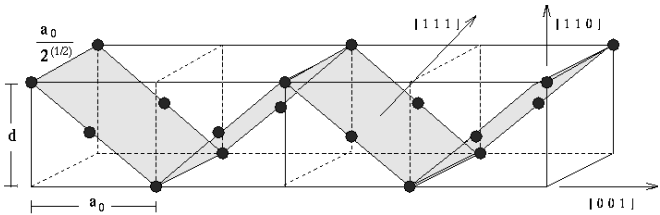}}
\caption{Faceting model geometry for the missing-row reconstruction:
the shaded area is the quasi-(111) reconstructed surface.
\label{cazzo4}}
\end{figure}

\paragraph*{The faceting model -- } The increase in workfunction and
surface stress with reconstruction  can be explained easily if we view
the reconstruction  as a microfaceting process into (111)-like facets
to vacuum. In previous works,\cite{garofalo,face}  faceting had
already been suggested as a key feature of the  missing-row
reconstruction: using our first-principle results, we now set up a
model  giving quantitative evidence in favor of the faceting
hypotesis. We then extend it to arbitrary $n$. The  reconstructed
surface, sketched  in Figure \ref{cazzo4}, has (111)-oriented facets
(shaded areas) at an angle of $\simeq \pm 35^{\circ}$ from the (110)
plane. The average coordination number and bonding geometry  of the
facet atoms are the same as on the (111) face, suggesting that the
facets may exhibit properties close to those of the (111) surface.   
We verify  such a similarity in a simple but stringent way, comparing
directly-calculated first-principles  energies and stresses for the
reconstructed surface  with  a model assuming the  energy and the
stress per unit facet area to be equal to those of the relaxed (111)
surface (also calculated ab initio), and integrating these values over
the facet area: 
\begin{equation}
\sigma^{\rm rec}_{2\times 1}={\sigma_{(111)}\over A_{(111)}^{1\times1}}\>
{a_0\>\sqrt{a_0^2+d^2} \over \sqrt{2}}\>.
\label{pippa}
\end{equation}
Here $a_0$ is the bulk lattice constant, $d$ the first-to-third plane
distance (see Figure \ref{cazzo4}), $\sigma_{(111)}/A_{(111)}^{1\times1}$
the energy per  unit area of the relaxed (111) surface. Analogously,
for the surface stress we take, 
\begin{equation}
\tau^{\rm rec}_y={\tau_{(111)}\over A_{(111)}^{1\times1}}\>
{a_0\>\sqrt{a_0^2+d^2} \over \sqrt{2}}\>,
\label{pippa2}
\end{equation} 
and for $\tau^{\rm rec}_x$ the projection of $\tau^{nr}_{(111)}$ along
$\hat{x}$.  For the unrelaxed surface, $d=a_0/\sqrt{2}$, and
$\sigma^{\rm rec}=2\>\sigma_{(111)}$. In Table \ref{t6} we compare the
ab initio  and model-derived values for the relaxed surfaces.  The
agreement of the two independent evaluations is indeed very good. It
is then plausible to identify the 2$\times$1 reconstructed (110)
surface with a spatially modulated  'quasi-(111)', quite similar to
the true (111) surface. The effect of outward and inward edges appears
to nearly compensate each other on average, and the reconstruction can
be thought of  as a transition from a rough surface (the
unreconstructed (110)) to a close-packed, high-coordination surface,
i.e. the quasi-(111) surface,  perturbed by  a relatively
long-wavelength corrugation ($\lambda$=2$a_0$). Thus, the increase of
workfunction is justified as it actually  reflects a transition from a
(110)-like to a  (111)-like situation (the behavior of the surface
stress is discussed later).   

The model easily explain the different behavior of Ir and Rh: the
energetics of reconstruction is governed by the ratio
$R=\sigma_{(110)}/\sigma_{(111)}$. From Equation (\ref{pippa}) we have
that the missing-row reconstruction occurs if 
\begin{equation}
{\sigma_{(110)}\over \sigma^{\rm rec}_{2\times 1}}
=R\,{A_{(111)}\over {a_0\over \sqrt{2}}\,\sqrt{a_0^2+d^2}}\equiv
\frac{R}{R_T}>1\>, 
\end{equation}
where $A_{(111)}=\sqrt{3}\,a_0^2/4$.
Therefore, the surface reconstructs if $R$ is larger than a
reconstruction threshold $R_T$ 
\begin{equation}
R_T\,=\,2\,\sqrt{2\over 3}\,\sqrt{1+\left({d\over a_0}\right)^2}
\simeq \,2\,\sqrt{1+{2\over 3}\left({\Delta d\over d^{\rm id}}\right)}\>,
\end{equation}
where $\Delta d$ is the change of $d$ from
its ideal value $d^{\rm id}=a_0/\sqrt{2}$, and the second equality is
valid to first order in ($\Delta d/d$).
$R_T$ is in other words the ratio of the facet area  (a  function of
$d$) to the (111) area. In Table \ref{t7} we report values of
$R$, $R_T$, and $\Delta d/d^{\rm id}$ for
Ir and Rh. If we disregard relaxations (i.e. put $\Delta d=0$), 
then $R_T$=2, and neither Ir nor Rh should reconstruct.
The reduction of $d$ (and hence of the facet area) is the essential
ingredient for Ir reconstruction. Rh, on the other hand,
 has nearly equal $R_T$, but a much smaller $R$, i.e. a much smaller
surface energy anisotropy.
Notice  however that Rh is actually just on the verge of reconstruction;
according to the model,  Rh would also reconstruct if it had 
 the same relaxations as Ir (i.e. if
the threshold values for Rh and Ir were the same).

\begin{figure}
\epsfclipon
\epsfxsize=8cm
\centerline{\epsffile{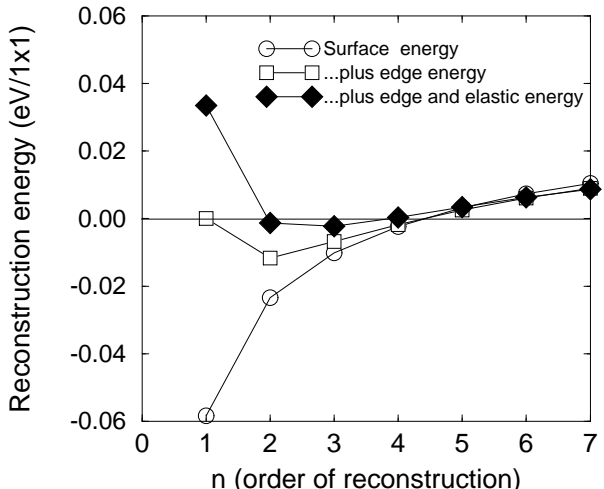}}
\narrowtext
\caption{The (110)--$n\times 1$ reconstruction energy
predicted by the faceting model as a function of $n$ (see text).}
\label{fig-faceting}
\end{figure}

\paragraph*{An $n$-dependent faceting model -- } 
Th model just discussed is $n$-independent. Since
 3$\times$1 faceted domains have been preferentially 
 observed in experiment, \cite{exp_ir}  we need to
  extend the model to  a general $n\times 1$
reconstruction. The vertical and horizontal 
distances between edges and valleys are now
 given by $d_n^{\rm ideal} = n a_0/2 \sqrt{2}$
and $d_{\rm horiz} = n a_0/2$. The surface energy (per 1$\times$1
area) generalizes to
\begin{equation}
\sigma^{\rm rec}_{n\times 1}=\sigma_{(111)}\,{4\over \sqrt{6}n}
\sqrt{n^2+\left({2 d_n \over a_0}\right)^2}
\end{equation}
where $d_n$ is the distance between first and (n+1)$^{\rm th}$ layer; 
the threshold ratio becomes
\begin{equation}
R_T(n)=
\frac{4}{\sqrt{6}\, n} \sqrt{n^2+
\left({n - \sum_i^n \rho_i\over \sqrt{2}}\right)^2}\>,
\end{equation}
with $\sum_i \rho_i=\delta d_n/d^{\rm ideal}_1$ the sum of 
relaxations expressed as fractions of  the ideal interplanar distance.
The reconstruction is favored for  
\begin{equation}
I = {\sigma^{\rm rec}_{n\times 1} -
\sigma_{(110)}\over \sigma_{(111)}} = R_T(n) - R <0\>.
\end{equation}
Since  the $\rho_i$'s are not known for layers far away from the  top,
we assume hypotetical relaxations  extrapolating to zero at  the tenth
plane below the surface  the values calculated\cite{mrs} for the
2$\times$1 case. $I\sigma_{(111)}$ is displayed as circles
in Figure \ref{fig-faceting}:  the reconstruction is favored for $n
\leq 4$. 
\noindent  The monotonic decrease of  $I\sigma_{(111)}$ as
 $n$$\rightarrow$\,1 disagrees with two facts:
{\it a)} the 1$\times$1 should be unstable with respect to the 2$\times$1
(although one may simply invoke the  inapplicability of the model for
 $n$=1); {\it b)}  experiments\cite{exp} show that  the 3$\times$1 is more
 stable than the 2$\times$1. To resolve  this, one has
 to  account for the contribution of the edges.   
Adopting a model due to Marchenko\cite{marcio} and
Shchukin {\it et al.}\cite{ciuco} based  on elasticity theory, we add
 an edge energy  and an   elastic energy to the reconstruction energy
 $I\sigma_{(111)}$. The total reconstruction energy
is given by 
\begin{equation}
\sigma^{\rm rec}_{n\times 1} = 
I \sigma_{(111)} - \frac{C_1}{n} + \frac{C_2}{n} {\rm ln} \frac{n}{2 \pi}.
\end{equation}
The parameter  $C_1$ is taken to be  equal to $I \sigma_{(111)}$, i.e.
 the energy needed to create on a (111) face the edges and valleys of a
(110)--$n\times$1. For  $C_2$, we use \cite{ciuco} 
\begin{equation}
C_2 = 8 (1 - \nu^2) \frac{\tau_{(111)}^2 \phi^2}{\pi Y a_0}\, ,
\label{eq-c2}
\end{equation}
with $\phi$=0.611 rad $\simeq$ 35$^{\circ}$
the faceting angle,
 $Y$=5.276 Mbar  the Young modulus,\cite{dati_ir} $\nu = 0.26$ the Poisson
ratio,\cite{dati_ir} and
 $a_0$=7.289 bohr the
calculated 
 lattice constant of Ir, $\tau_{(111)}=1.96$
eV/(1$\times$1) the first-principles calculated surface stress of Ir
(111).
The successive approximations are depicted in Figure \ref{fig-faceting}.
The circles are $I \sigma_{(111)}$; adding the
edge energy we obtain (squares) a minimum at $n$=2, and stable $n$=3
and $n$=4; the complete expression (filled diamonds)
 gives $n$=2,3 stable with $n$=3 
lowest in energy. The calculated reconstruction 
energy is not reproduced quantitatively, but its qualitative
behavior is now correct.

\begin{figure}
\epsfclipon
\epsfxsize=8cm
\centerline{\epsffile{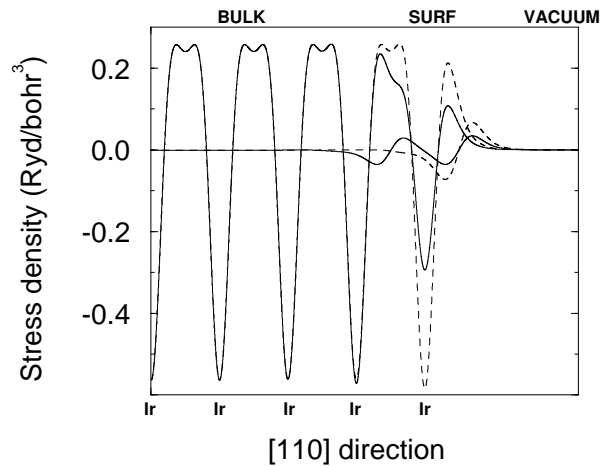}}
\narrowtext
\caption{Planar and macroscopic averages of stress density for the
reconstructed (full lines) and unreconstructed
(dashed lines) Ir (110) surface.
\label{cazzo10}}
\end{figure}

\paragraph*{Explaining the stress increase: application of the stress
density -- }  
In the light of the faceting model, the rationale 
for the increase in tensile stress is that reconstruction, while
removing rows from the surface, exposes new ones to the vacuum,
causing the   stress stored in the surface rows atoms to be
spatially redistributed, and specifically, transferred towards
 the substrate. To support this argument directly, we introduced,
\cite{long}  
in analogy to the Chetty-Martin
 energy density,\cite{cm} the stress density, defined as a tensor 
${\cal T}_{\alpha\beta}({\bf r})$, functional of the charge density, 
whose integral over the cell equals the macroscopic stress:
\begin{equation}
\int_{\Omega} d{\bf r}\,{\cal T}_{\alpha\beta}({\bf r})=
\tau_{\alpha\beta}\>.
\label{prima}
\end{equation}
The detailed formulation, as well as a discussion of gauge dependence
issues, will be given elsewhere.\cite{long} The core idea is that 
while surface   stress only provides information on certain phenomena  at
surfaces in an integrated fashion, the stress density allows to
analyze {\it locally} its different components. In the present case,
we need to consider the  planar  average  
$\overline{\cal T}_{\alpha\beta}(z)$ of the stress density along
the [110] 
direction, and the average of
$\overline{\cal T}_{\alpha\beta}(z)$  over the inter-layer distance  (the
macroscopic average\cite{bbr}). In Figure \ref{cazzo10},
 we compare  planar  and macroscopic average of the stress density of 
\noindent the unreconstructed 
(dashed lines) and reconstructed (full lines) surfaces. 
Both surfaces are considered in their
unrelaxed configuration for graphical convenience and,
exploiting  inversion symmetry, we only display one slab half.
Note  that the two inequivalent stress components along
$\hat{x}$ and $\hat{y}$ are indistinguishable on this scale.

The macroscopic averages 
 vanish into bulk (by the  equilibrium condition
of vanishing stress) and in the vacuum,
and they deviate  from zero whenever surface effects arise. Thus, 
they enable one to establish the effective boundaries of the surface  
region as far as stress propagation is concerned. 
The planar averages show periodic oscillations in the bulk
region, and  vanish into the vacuum. The negative peaks 
are tensile contributions arising mostly from
 electron-ion interactions, and are located at the
 atomic  positions. The positive peaks rising between
atomic planes store the compressive stress of
essentialy kinetic origin. 

The stress transfer upon reconstruction is easily detected
moving from  left to right in Figure \ref{cazzo10}:
 for the unreconstructed surface (dashed line), a non-zero
macroscopically-averaged stress density is only visible in the first
layer, whereas for the reconstructed one it starts off to non-zero
values one layer earlier. The stress
is indeed redistributed from the surface to the substrate.    Focusing
on the planar averages,  the   
 negative (tensile) peak  in the first layer of the reconstructed
surface is  reduced with  respect to the unreconstructed one.
This corresponds to the expected reduction of
tensile stress upon removal of  the [1$\overline{1}$0]-oriented 
atomic rows.  On the other hand, the positive peak between surface and
substrate is also strongly reduced, and it compensates (in fact,
overcompensates) the reduction of the tensile component. In other
words,  row removal reduces  the tensile stress in the surface plane,
but at the same time it allows electrons to delocalize into the
troughs thus created,   reducing the compressive kinetic stress.   
Ho and Bohen\cite{hb} arrive at essentially the same conclusion by
analyzing the individual contributions to the total energy, and
finding a reduction in electron kinetic energy, larger in Au
than in Ag. 

In summary, we showed that the (110) missing-row reconstruction can be 
understood as microfaceting  of the  (110) into a bent, quasi-(111)
surface. The reconstruction reduces the effective surface roughness
 and causes a stress transfer from
the surface towards the substrate. The latter effect has been
analyzed by  means of the stress density. The different
behavior between $5d$ and $4d$ elements can be traced back to the
greater (110)/(111) surface energy anisotropy, and ultimately to
the relativistic character of the heavier elements.\cite{fms} 

Support from CRS4 Cagliari in the form of computing time on its
IBM SP2 is acknowledged.
V.F.'s stay at WSI was supported by the Alexander von Humboldt-Stiftung.

\vspace{-0.4cm}
\begin{table}
\refstepcounter{table}
\parbox{\hsize}{TABLE~\ref{t5}. 
Surface energy $\sigma^{\rm unrec}$, stresses  $\tau_x^{\rm unrec}$ and
$\tau_y^{\rm unrec}$,
and workfunction $W^{\rm unrec}$  of unreconstructed  (110); 
$H^{rec}$, $\Delta W$, $\Delta\tau_x$ and $\Delta\tau_y$ are changes
in  previous quantities upon reconstruction.
Results  in eV or eV/atom.} 
\label{t5}
\begin{tabular}{ccccccccc}
        &  $\sigma^{\rm unrec}$&$H^{\rm rec}$&$W^{\rm unrec}$&$\Delta W$&$\tau_x^{\rm unrec}$&
$\Delta\tau_x$& $\tau_y^{\rm unrec}$&$\Delta\tau_y$ \\
\hline
 Ir     & 2.59&--0.03& 5.42&+0.15&--1.70&--0.44&--3.21&--0.50 \\
\hline
 Rh     & 1.89& 0.07& 5.07& +0.15& --1.25& --0.09 &--2.01&--0.13\\
\end{tabular}
\end{table}
\vspace{-0.6cm}
\begin{table}
\refstepcounter{table}
\parbox{\hsize}{TABLE~\ref{t6}.
Reconstruction energies $H^{\rm rec}$ and reconstructed surface
stress $\tau^r$ (eV/atom) as calculated directly (c) and
 by the faceting model (m) (see text).}
\label{t6}
\begin{tabular}{lrrrrrr}
 & \multicolumn{3}{c}{Ir} & \multicolumn{3}{c}{Rh}  \\
\hline\hline
 & $H^{\rm rec}$ &  $\tau_x^{\rm surf}$  & $\tau_y^{\rm surf}$ &
$H^{\rm rec}$ &  $\tau_x^{\rm surf}$  & $\tau_y^{\rm surf}$ \\
\hline\hline
 c &--0.03 & --2.14 &--3.71 &0.07 &--1.34 &--2.14 \\
\hline
 m &--0.04 & --2.20 &--3.82 &0.03 &--1.32 &--2.28 \\
\end{tabular}
\end{table}
\vspace{-0.6cm}
\begin{table}
\refstepcounter{table}
\parbox{\hsize}{TABLE~\ref{t7}.
$R=\sigma_{[110]}/\sigma_{[111]}$,
 faceting  threshold $R_T$;  
relaxation change of first-third layer
distance $\Delta d/d^{\rm id}$.}
\label{t7}
\begin{tabular}{lccc}
      & $R$ & $R_T$ & $\Delta d/d^{\rm id}$  \\
\hline\hline
  Ir  &  1.977 & 1.947 & --0.078\\
\hline
  Rh  & 1.948 & 1.953 &--0.069 \\
\end{tabular}
\end{table}

\end{multicols}
\end{document}